# HST IMAGES OF NEARBY LUMINOUS QUASARS[1]

John N. Bahcall, Sofia Kirhakos,

Institute for Advanced Study, School of Natural Sciences, Princeton, NJ 08540

and

Donald P. Schneider

Institute for Advanced Study, School of Natural Sciences, Princeton, NJ 08540

and

Department of Astronomy and Astrophysics, The Pennsylvania State University, University Park, PA 16802

## ABSTRACT

Strong upper limits are placed on the visual-band brightnesses of galactic hosts for four luminous, radio-quiet quasars with redshifts between 0.16 and 0.24 that were studied with the HST's Wide Field/Planetary Camera-2. Typical upper limits on the luminosities of galactic hosts are about 1.4 mag fainter than $L^*$ for spirals and about 0.5 mag fainter than $L^*$ for ellipticals. The galactic hosts of the quasars are more than a magnitude and a half fainter than the median integrated absolute magnitude of Seyfert galaxies. If the detection limits are determined using featureless simulated galaxies instead of observed galaxy images, then the detection limits for spirals are 0.5–1.0 mag less stringent. These results are consistent with the hypothesis that the quasar phenomenon corresponds to the early stages of galaxy formation, before extensive star formation occurs.

*Subject headings:* quasars: individual (PG 0953+414, PG 1116+215, PG 1202+281(GQ COM), PG 1307+085)

---

## 1. INTRODUCTION

The imaging of low-redshift quasars at high angular resolution ($\sim 0.1''$) is one of the principal scientific goals for which the Hubble Space Telescope was designed. It has long been recognized that quasars might reside in galaxies and that the quasar phenomenon might be strongly influenced by the presence of companion galaxies. There is an extensive observational literature on this subject; some representative papers are Kristian (1973), Heckman et al. (1984), Boroson, Persson, & Oke (1985), Stockton & MacKenty (1987), Yee (1987), Hutchings, Janson, & Neff (1989), and Dunlop et al. (1993). The extraordinary optical characteristics of the repaired HST make possible improved observational studies of the hosts of small-redshift quasars. We present in this letter the initial results that have been obtained for the first four objects in a sample of 18 luminous ($M_V < -22.9$, for $H_0 = 100$ km$^{-1}$s$^{-1}$Mpc$^{-1}$, $\Omega_0 = 1.0$) nearby ($z < 0.30$) radio-quiet and radio-loud quasars selected randomly from the Veron-Cetty & Veron (1993) catalog.

Our principal results for the quasars and their hosts are summarized in Table 1, which lists the following quantities for each of the four quasars: the date observed, the redshift and the apparent $V$-magnitude (from Veron-Cetty & Veron 1993), the absolute $V$-magnitude, and the range of galaxy magnitudes at which six representative galaxies (selected from HST observations) could be detected when artificially placed underneath the quasar images. The last column gives the average limiting absolute magnitude at which the six representative galaxies could have been detected as hosts. Accurate ($1''$) coordinates and multicolor CCD photometry for three of the four quasars are given in Kirhakos et al. (1994). All four of the quasars are radio quiet (Kellermann et al. 1989).

## 2. OBSERVATIONS

The quasars were observed with the Wide Field/Planetary Camera-2 (WFPC2) through the F606W filter (see Burrows 1994 for a description of the instrument and the response curve of the filter). The F606W filter is similar to the $V$ bandpass but is slightly redder; the mean wavelength and FWHM of the F606W system response are 5940 Å and 1500 Å, respectively. This filter was chosen because of its high throughput, which was the primary consideration for the present exploratory program. The STScI photometric calibration was verified by comparing, in quasar fields obtained at Palomar by Kirhakos et al. (1994), HST galaxy magnitudes with ground-based photometry of the same galaxies. For the purposes of this paper, the $V$ and the $F606W$ photometric bands are sufficiently similar (Bahcall et al. 1994; J. Holtzman 1994, private communication) that we will sometimes use $V$ and $m_{F606W}$ interchangeably.

All quasar images were located at a distance of about $6'' \pm 1''$ from the center of Wide Field CCD 3 and, except for PG 1307+085, were observed for three separate exposures of 1100 s, 600 s, and 100 s; the exposures for PG 1307+085 lasted 1400 s, 500 s, and 200 s. The image scale of the Wide Field detectors is $0.10''$ pixel$^{-1}$, which means that the data are severely undersampled. The quasar images are saturated in all of the exposures. Typically, there were about 9 saturated pixels in the shortest exposures and 84 saturated pixels in the longest exposures.

The point spread function (PSF) was determined using four saturated exposures (12s duration each) of an isolated $V$ = 10.5 mag star (F141) in M67 that was placed on the corners of a box $3''$ on a side. The box was centered at a point displaced $2.5''$ away from the central pixel of Wide Field CCD 3 in each of the two orthogonal directions. In addition, a 2.6s and a 70s exposure were also obtained at one corner of the box. The M67 observations, which were obtained on June 5, 1994, cover the saturation range present in the quasar images.

The initial data processing (bias frame removal and flat-field calibration) was performed at the Space Telescope Science Institute with their standard software package. The individual images of each quasar were aligned to better than 0.3 pixel; this made it easy to identify and eliminate cosmic ray events. The flat-field corrections were based upon pre-flight calibrations; these calibrations remove the small-scale (few pixel) sensitivity variations. However, there are apparent large-scale sensitivity variations present in the calibrated data. The typical signal and rms of the noise of the sky in the long exposures (in detected photons) are 90 and 14, respectively. The sky level corresponds to a surface brightness of approximately 22.4 mag sq arc sec. The observed noise is slightly larger than expected from a combination of shot noise and CCD readout noise (7 electrons per pixel). The formal detection limits for extended objects, calculated from these numbers, are extremely faint. However, one limiting factor for the detection of galaxies is the imperfect match between the data and the pre-flight flat-field calibrations.

It is difficult to measure magnitudes for saturated stellar images, but we determined–as a further check on our photometry–approximate magnitudes for the quasars. We estimated the magnitudes by fitting to the quasar images in the shortest exposures a scaled PSF of the 2.6s exposure of the M67 calibration star F141 and also by fitting a theoretical model (Burrows 1994) of the F606W PSF. The results are in satisfactory agreement with the magnitudes given in the Veron-Cetty & Veron (1993) catalog.

## 3. HOST GALAXIES

The four luminous quasars described in this paper do not occur in bright galaxies. We established this result by artificially placing a series of observed and simulated galaxies underneath the quasar images, and determining, with both objective and subjective (visual inspection of the data) techniques, the surface brightness levels at which the galaxies are detectable.

In the analysis that follows, we quote the results obtained using the different stellar PSFs determined from the M67 calibration data. In order to test the robustness of our conclusions, we also used: 1) a PSF determined from bright stars not located at the center of the WFC-3 CCD, 2) an empirical PSF formed by using our image of PG 1116+215, (This is the brightest quasar in our sample.) and 3) a theoretical F606W PSF calculated by Burrows (1994). All of the PSFs gave essentially the same answers. Double-blind experiments were also performed in which one of us placed galaxies of different morphological types and brightnesses underneath quasar images and the other two workers visually inspected the images, with a best-fit stellar PSF subtracted, to determine if galaxies were identifiable.

The first stage of the analysis consists of subtracting a stellar PSF from the observed quasar image. The best fit to the quasar image was determined using the downhill simplex method (described in Press et al. 1986) using the $\chi^2$ calculated in an annulus with inner and outer radii of $\approx 0.5''$ and $2.0''$; the three parameters were the location (two coordinates) and the peak brightness of the PSF. Pixels that were saturated in either the quasar data or the PSF data, or that showed a large disagreement between the PSF and the quasar data ($3\sigma$ from the mean deviation), were eliminated from the $\chi^2$ fitting.

Figure 1 shows the result of subtracting a best-fit stellar image (of M67 calibration star F141) from the long-exposure images of each of the four quasars. Each panel is $20'' \times 19''$. The contrast has been set to emphasize low-surface brightness features; in the long exposures, the quasar images are saturated to a radius of $\approx 0.3''$.

The best case for a candidate host galaxy is provided by the image of PG 1116+215. There appears to be a faint (23.4 mag/sq. arc s), smooth ring-like structure with a radius of about $2''$ surrounding almost half of PG 1116+215, plus an additional diffuse protrusion. If this nebulosity is real, it corresponds to a host galaxy of $m_{F606W} \approx 19.0$ (0.1 mag fainter than the average limiting magnitude estimated from the simulations and listed in Table 1), and a total diameter of $\sim 20$ kpc. We are somewhat suspicious of this candidate nebulosity because of its smooth, symmetric shape.

PG 1307+085 also shows evidence of non-stellar morphology, which after close examination we believe to be mostly due to a doughnut-shaped optical ghost. This faint, fuzzy extension, which is seen in the subtracted image of PG 1307+085(Figure 1d), at $PA \approx 92°$, has a total $V$ magnitude of $\approx 22.4$ mag. Somewhat similar features are recognizable as ghosts of bright stars in archival WFPC2 data. Calibration stars placed close to the center of WF3, with approximately the same saturation in F606W as PG 1307+085, were required to recognize the ghost-character of this apparent nebulosity.

In order to quantify our detection limits for faint nebulosity, we created a set of images of individual galaxies that were added to the HST quasar images. We used simulated test galaxies (both exponential disks and de Vaucouleurs spheroids) as well as images of a number of galaxies observed with the Wide Field CCDs of the WFPC2 through the F606W filter. The limits set on the observed galaxies are stronger, typically by 0.5–1.0 mag, than the limits set on the artificial galaxies, because the artificial galaxies have smooth profiles that are difficult to distinguish from imperfect flat-field corrections.

Figure 2 shows the collection of six observed galaxy images whose surface brightnesses were scaled in order to determine empirically our sensitivity to extended nebulosity centered on the quasar. At the redshifts of the quasars, the sizes of the galaxies, $1''$ to $4''$, correspond to approximately $2 - 8$ kpc (in terms of effective radii or scale lengths, depending on the individual profiles). All of the galaxies were taken from HST observations in this program; they represent a variety of galaxy types. Fig. 2a represents a pair of interacting spiral galaxies, Fig. 2b is an edge-on spiral, Fig. 2c is a ringed spiral seen face on, Fig. 2d is a smooth elliptical, Fig. 2e is an S0 or an Sa, and Fig. 2f appears to be a disturbed barred spiral.

To determine the detection limits for a given galaxy, the galaxy was added to a quasar observation and the $\chi^2$ for the best PSF fit (for each assumed PSF) was recorded as a function of galaxy brightness (in all cases the center of the galaxy and the quasar were aligned). The galaxy sizes were held fixed as they appear in Fig. 2. As the galaxy image was made progressively fainter, the $\chi^2$ dropped until the galaxy signal became much lower than the background noise. When the computed $\chi^2$ equaled twice the asymptotic value, a clear break point in the $\chi^2$ vs brightness curves, we were confident that we could detect the galaxy. Visual inspection of the simulated images showed that in some cases (in particular for the edge-on spiral, Fig. 2b), the galaxy could be detected $\approx 1$ mag fainter than indicated by the $\chi^2$ test described above.

Table 2 gives the limiting magnitude down to which each of the six galaxies shown in Fig. 2 could be detected when the galaxy was centered on the quasar image. The magnitude range of the faintest host galaxies that could be detected are summarized in the next to last column of Table 1. The faintest galaxies, $m_{\rm lim} \sim 20$ mag, could be detected when the galaxy had the form of Fig. 2b, the edge-on spiral. The most difficult morphology to detect was the featureless elliptical galaxy in Fig. 2d, for which $m_{\rm lim} \sim 18.5$ mag. For each quasar, the limits of detectability of the galaxy images spanned a range of 1 to 2 mag (see Table 2), which the average being about four magnitudes fainter than the quasar.

The galaxies could be detected to slightly fainter magnitudes if the quasar and galaxy centers were not coincident. As a numerical example, we displaced the centers by 0.3" ($\approx 0.6$ kpc) and repeated the sensitivity experiments described above. On average, the four spirals shown in Fig. 2 could be detected to 0.1 mag fainter limits than indicated in Table 2. The increase in detection sensitivity was 0.2 mag on average for the two ellipticals.

For each quasar, we averaged the limiting magnitudes given in Table 2 and used this value to compute an average absolute magnitude of a detectable host galaxy (see the last column of Table 1). The galaxies shown in Fig. 2 could, on average, be detected to 4 mag fainter than the quasars. The typical magnitude limits are $M_{\rm F606W}({\rm lim}) = -19.4$ mag, which is about 1.1 mag fainter than $L^*$ for field galaxies (cf. footnote a of Table 1).

Could we have missed host galaxies that were very small in size? We tested this possibility by using an observed spiral galaxy with a total size of only 2" ($\sim 4$ kpc at the quasar distances). We were able to detect this spiral galaxy to a limiting magnitude as faint, or slightly fainter, than the limits given in column (6) of Table 2. We conclude that our images were sensitive to even relatively small galaxies.

It is difficult to compare our results with previous ground-based observations because of the greater resolution available in the HST images. Consider, for example, the problem of distinguishing between light from galactic companions (galaxies physically distinct from the quasar but lying nearby on the sky) and light from galactic hosts (galaxies in which the quasar phenomenon occurs). HST observations determine more easily than terrestrial observations whether observed light originates from companions or from hosts One of our objects, PG 0953+414, has been investigated with ground-based K-band images by Dunlop et al. (1993) and B-band and R-band images by Hutchings et al. (1989). Dunlop et al. (1993) report the presence of a companion galaxy close to PG 0953+414; four companion objects can be seen projected close to PG 0953+414 in Figure 1a , the brightest of which is a galaxy at $PA \approx 330°$ with $m_{F606W}$ = 22.9 mag (The other three objects are a star at $PA \approx 320°$, $m_{F606W}$ = 24.2 mag; a galaxy at $PA \approx 5°$, $m_{F606W}$ = 24.6 mag; and a galaxy at $PA \approx 175°$, $m_{F606W}$ = 24.3 mag.). Hutchings et al. (1989) have suggested that PG 0953+414 might have a large (56 kpc), low surface brightness host, so faint that we would not have detected it. We do, however, see for PG 0953+414 what may be an extended, low surface brightness feature ($\mu_V$ ∼ 24 mag per square arcsecond, between $PA \approx 100°$ to ∼ $PA \approx 180°$ ) at ∼ 3″ to ∼ 5″ from the quasar image.

## 4. SUMMARY

The principal result of this paper is that none of the four quasars we have studied resides in a luminous galaxy (see Table 2). If the sample of galaxies shown in Fig. 2 is representative of the average host, our detection limits indicate that the host galaxies of low-redshift, radio-quiet quasars are fainter than $L^*$ galaxies.

Many authors have suggested that radio-quiet quasars reside in luminous spiral galaxies. We would be able to detect the four spirals shown in Fig. 2 to an average of $M_V^*$ = −19.1 mag, corresponding to a limit that is about 1.4 magnitude fainter than an $L^*$ galaxy ($M_V^*$ = −20.5 mag, cf. Kirshner et al. 1983, Esthasiou, Ellis, and Peterson 1988).

We could detect the two ellipticals shown in Fig. 2 to an average of $M_V^*$ = −20.0 mag, which is about two magnitudes fainter than the total magnitudes of the brightest elliptical galaxies in rich clusters (cf. Hoessel and Schneider 1985). The detection limits we place ignore k-corrections. If the hosts are as red as a giant elliptical, then the limits would be approximately 0.5 mag brighter.

Our results are surprising given the well-established continuity between Seyfert galaxies, other active galactic nuclei, and quasars. The median integrated absolute magnitude of classical Seyfert galaxies is, for example, about $M_V = -21.0$ mag (Osterbrock and Martel 1993, Huchra and Burg 1992). The host galaxies for the quasars studied in this paper are typically fainter than $M_V = -19.4$ mag , more than a magnitude and a half fainter than the median Seyfert galaxy. The results presented here do not fit easily into the standard paradigm of quasars being fueled by gas and stars from a surrounding, well-developed galaxy. One possible explanation of our observations is that, if the quasars studied in this paper are typical, the luminous quasar phenomenon may signal the early stages of galaxy formation before extensive star formation has occured. Of course, some stars must be present even in this stage to produce the heavy-elements that are observed in quasar spectra. According to this "quasar first, galaxy second" scenario, the less luminous stages of the AGN phenomenon would occur after extensive star formation took place and the original quasar dimmed. If, as is commonly believed, most galaxies were formed at moderate to large redshifts, then the "quasar first" scenario would also imply, as observed, that quasars are more numerous at earlier cosmological epochs than they are today.

**Note Added in Proof.** Limits on the luminosities of host galaxies depend on the morphology, scale lengths, and (in some cases) the orientation of the presumed host galaxy. There are combinations of morphology, scale length, and orientation that, in principle, would permit host galaxies brighter than $L^*$ to have escaped detection in the analysis reported here. The authors are currently preparing a paper that describes in greater detail the analysis of the images of the four quasars discussed in the present paper and the images of four additional quasars obtained subsequently. In this next paper, we discuss numerous checks that we have made that are designed to determine how well we would have detected host galaxies that have been reported by other observers.

We are grateful to C. Burrows, J. Felten, C. Flynn, R. Gilliland, P. Goldreich, A. Gould, J. Holtzman, B. Jannuzi, J. Ostriker, J. MacKenty, D. Osterbrock, P. Schechter, and M. Strauss for valuable discussions; and to David Saxe who wrote much of the software used to prepare the figures. We would like to thank Digital Equipment Corporation for providing the DEC4000 AXP Model 610 system used for the computationally-intensive parts of this project. This work was supported in part by NASA contract NAG5-1618 and grant number GO-2424.01 from the Space Telescope Science Institute, which is operated by the Association of Universities for Research in Astronomy, Incorporated, under NASA contract NAS5-26555.

# REFERENCES

Bahcall, J. N. , Flynn, C., Gould, A., and Kirhakos, S. 1994 ApJ Lett. (this issue)

Boroson, T. A., Persson, S. E., & Oke, J. B. 1985, ApJ, 293, 120

Burrows, C. J. 1994, Hubble Space Telescope Wide Field and Planetary Camera 2 Instrument Handbook, Version 2.0 (Baltimore: STScI)

Burrows, C. J. 1994 (private communication)

Dunlop, J. S., Taylor, G. L., Hughes, D. H., & Robson, E. I. 1993, MNRAS, 264, 455

Esthasiou, G., Ellis, R. S., & Peterson, B. A. 1988, MNRAS, 232, 431

Heckman, T. M., Bothum, G. D., Balick, E., & Smith, E. P. 1984, AJ, 89, 958

Hoessel, J. G., & Schneider, D. P. 1985, AJ, 90, 1648

Huchra, J. & Burg, R. 1992, ApJ, 393, 90

Hutchings, J. B., Janson, T., & Neff, S. G. 1989, ApJ, 342, 660

Kellermann, K. I. , Sramek, R., Schmidt, M., Shaffer, D. B., & Green, R. 1989, AJ, 98, 1195

Kirhakos, S., Sargent, W. L. W., Schneider, D. P., Bahcall, J. N., Jannuzi, B. T. Maoz, D., & Small, T. A. 1994, PASP, 106, 646

Kirshner, R. F., Oemler, A., Schechter, P. L., & Shectman, S. A. 1983, AJ, 88, 1285

Kristian, J. 1973, ApJ, 179, L61

Osterbrock, D. E. & Martel, A. 1993, ApJ 414, 552.

Press, W. H., Flannery, B. P., Teukolsky, S. A., & Vetterling, W. T. 1986, Numerical Recipes: The Art of Scientific Computing (Cambridge: Cambridge University Press), 289

Schechter, P. 1976, ApJ, 203, 297

Stockton, A. & MacKenty, J. W. 1987, ApJ, 316, 584

Veron-Cetty, M. P., & Veron, P. 1993, A Catalogue of Quasars and Active Nuclei (Sixth Edition), ESO Scientific Report, No. 13

Yee, H. K. C. 1987, AJ, 94, 1461

---

Table 1: Sample of Quasars and Detectable Hosts

| Object | Date 1994 | $z$ | $V$ | $M_V(\mathrm{QSO})^a$ | $m_{\mathrm{lim}}(\mathrm{host})$ F606W | $\langle M_{\mathrm{lim}}(\mathrm{host})\rangle^a$ F606W |
|---|---|---|---|---|---|---|
| PG 0953+414 | 3 Feb | 0.239 | 15.3 | −24.1 | $\gtrsim 19.1 - 20.4$ | −19.5 |
| PG 1116+215 | 8 Feb | 0.177 | 15.0 | −23.7 | $\gtrsim 18.5 - 19.7$ | −19.8$^c$ |
| PG 1202+281$^b$ | 8 Feb | 0.165 | 15.6 | −23.0 | $\gtrsim 18.5 - 20.3$ | −19.0 |
| PG 1307+085 | 4 Apr | 0.155 | 15.3 | −23.1 | $\gtrsim 18.4 - 20.2$ | −19.1 |

$^a$ Computed for $\Omega_0 = 1.0$ and $H_0 = 100$ km s$^{-1}$Mpc$^{-1}$. In this cosmology, brightest cluster galaxies have $M_V \approx -22.0$ mag (Hoessel and Schneider 1985) and the characteristic Schechter (1976) magnitude for field galaxies is $M_V^* = -20.5$ mag (Kirshner et al. 1983; Esthasiou, Ellis, & Peterson 1988).
$^b$ GQ Com
$^c$ Candidate galaxy has $M_V = -19.7$.

Table 2: Limiting Magnitudes at Which Galaxies in Fig. 2 Could Be Detected as Hosts

| Quasar | kpc/1″ | 2a | 2b | 2c | 2d | 2e | 2f | $\langle m_{\mathrm{lim}}(\mathrm{host})\rangle$ |
|---|---|---|---|---|---|---|---|---|
| PG 0953+414 | 2.4 | 20.1 | 20.4 | 19.6 | 19.1 | 19.7 | 20.2 | 19.9 |
| PG 1116+215 | 1.9 | 18.9 | 19.7 | 18.9 | 18.5 | 18.6 | 19.0 | 18.9 |
| PG 1202+281 | 1.8 | 20.3 | 20.3 | 19.4 | 18.5 | 19.0 | 19.9 | 19.6 |
| PG 1307+085 | 1.8 | 20.1 | 20.2 | 18.5 | 18.4 | 18.6 | 19.7 | 19.3 |

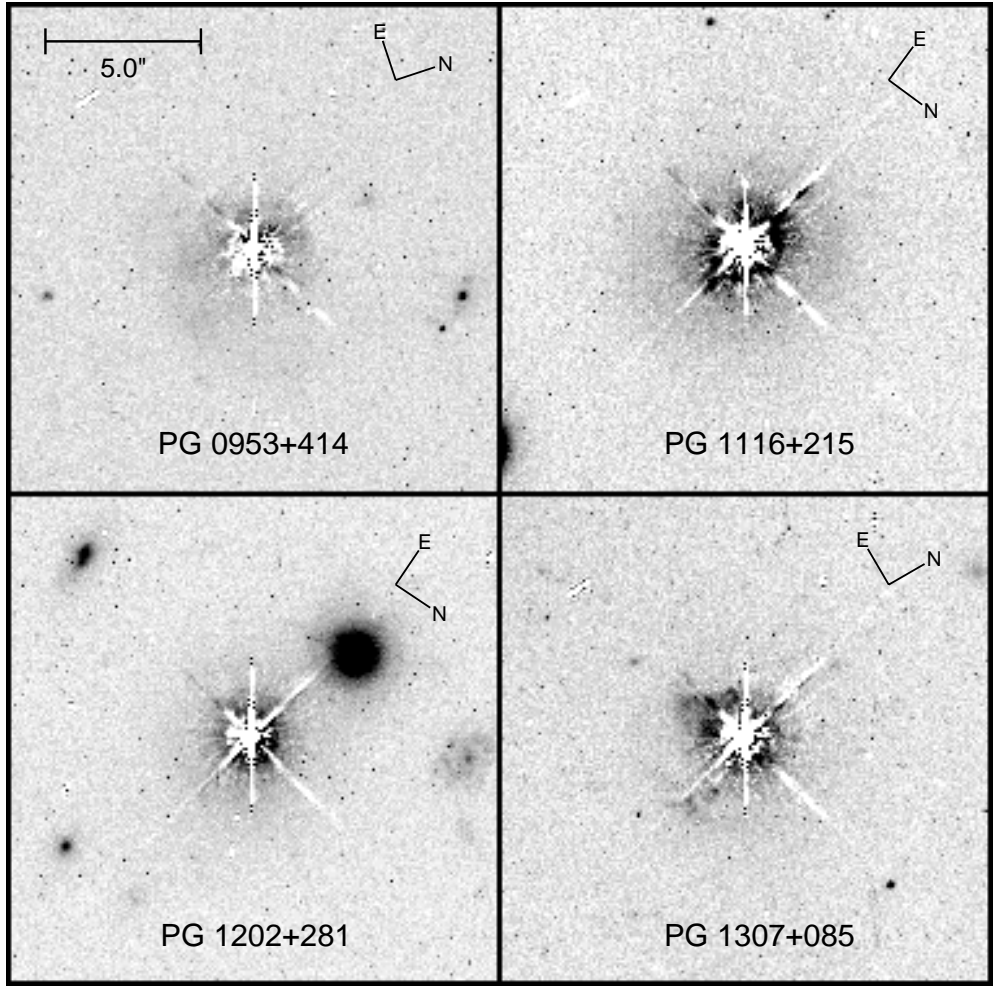

Fig. 1.— HST images of four nearby quasars. This figure shows the "long" (1400 s for PG 1307+085; 1100 s for the others) F606W observation of the four quasars discussed in this paper. A best-fit stellar image from the M67 calibration data has been subtracted from each quasar image. Each of the panels is $20'' \times 19''$; the image scale is $0.10''$ pixel$^{-1}$. Here $1'' \approx 2$ kpc $(0.2/z)$.

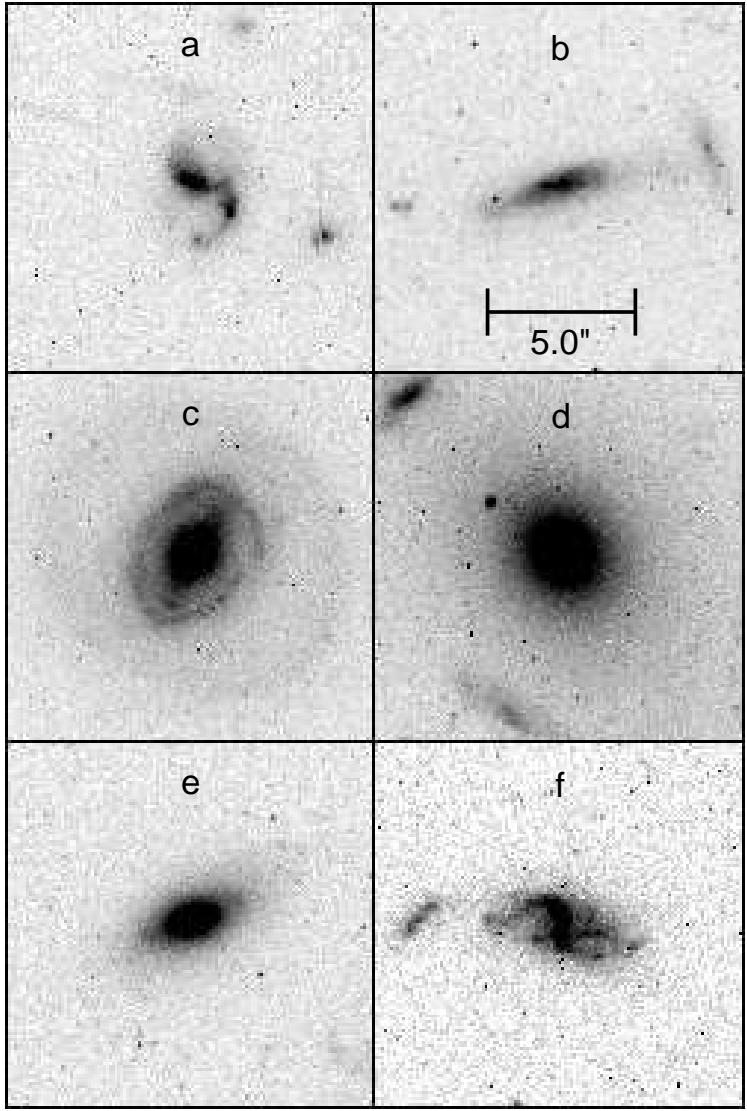

Fig. 2.— This figure shows the six galaxies, taken from the fields of the quasar HST observations, that were used to determine the sensitivity of the HST images to different types of quasar host galaxies. Each panel is 12.5″ on a side.